\documentclass[a4paper,fleqn]{cas-sc}
\usepackage{diagbox}
\usepackage{lipsum}
\usepackage{caption}
\usepackage{subcaption}
\DeclareUnicodeCharacter{1ED3}{\~o}
\usepackage[authoryear,longnamesfirst]{natbib}
\usepackage{wrapfig}
\usepackage[ruled,vlined]{algorithm2e}
\usepackage{algpseudocode} 
\usepackage{multirow}
\graphicspath{ {Figures/} }
\def\tsc#1{\csdef{#1}{\textsc{\lowercase{#1}}\xspace}}
\tsc{WGM}
\tsc{QE}
\tsc{EP}
\tsc{PMS}
\tsc{BEC}
\tsc{DE}

\begin{document}
\let\WriteBookmarks\relax
\def\floatpagepagefraction{1}
\def\textpagefraction{.001}
\shorttitle{Two Decades of Scientific Misconduct in India}

\title [mode = title]{Two Decades of Scientific Misconduct in India: Retraction Reasons and Journal Quality among Inter-country and Intra-country Institutional Collaboration}                      

\author[1]{Kiran Sharma\corref{cor1}}
\ead{kiransharma1187@gmail.com}

\address[1]{School of Engineering and Technology, BML Munjal University, Gurugram, Haryana-122413, India}
\cortext[cor1]{Corresponding author}

\begin{abstract}
Research stands as a pivotal factor in propelling the progress of any nation forward. However, if tainted by misconduct, it poses a significant threat to the nation's development. Recognizing the importance of genuine research and understanding the ramifications of scientific malpractice are essential for grasping a nation's trajectory of growth. This study aims to scrutinize various cases of deliberate scientific misconduct by Indian researchers. A comprehensive analysis was conducted on 3,244 retracted publications sourced from the Retraction Watch database. The upward trend in retractions is alarming, although the decreasing duration of retractions indicates proactive measures by journals against misconduct. Approximately 60\% of retractions stem from private institutions, with fake peer reviews identified as the primary cause of misconduct. This trend could be attributed to incentivizing publication quantity over quality in private institutions, potentially fostering unfair publishing practices. Retractions due to data integrity issues are predominantly observed in public and medical institutions, while retractions due to plagiarism occur in conference proceedings and non-Scopus-indexed journals. Examining retractions resulting from institutional collaborations reveals that 80\% originate from within the country, with the remaining 20\% being international collaborations. Among inter-country collaborations, one-third of retractions come from the top two journal quartiles, whereas, in intra-country collaborations, half of the retractions stem from Q1 and Q2 journals. In intra-country collaborations, 86\% of retractions are from articles and conference proceedings, while in inter-country collaborations, 87\% are solely from articles. Clinical studies retracted from intra-country collaborations are mostly from Q3 and Q4 journals, whereas in inter-country collaborations, they primarily come from Q1 journals. Regarding top journals by the number of retractions in intra-country collaborations, they belong to the Q2 and Q4 categories, whereas in inter-country collaborations, they are in Q1. This underscores the fact that retractions involving international collaborations tend to occur in high-quality journals, whereas those involving domestic collaborations are more commonly associated with journals of moderate to low quality.
\end{abstract}


\begin{highlights}
\item The increasing number of retractions is concerning, while the decreasing duration of retractions suggests that journals are taking active measures against misconduct.

\item Approximately 60\% of retractions stem from private institutions.  Fake peer review is recognized as the primary cause of scientific misconduct in Indian research carried out by private institutions and colleges.

\item Retractions due to data integrity predominantly occur in public and medical institutions, while retractions due to plagiarism occur in journals not indexed in Scopus and conference proceedings.

\item Institutional collaborations indicate that 80\% of retractions originate domestically, while the remaining 20\% stem from international collaborations.

\item In inter-country collaborations, one-third of the total retractions originate from the top two journal quartiles, whereas in intra-country collaborations half of the total retractions originate from the top two journal quartiles, i.e. Q1 and Q2.

\item In intra-country collaborations, the top journals, based on the number of retractions, fall within the Q2 and Q4 categories, whereas in inter-country collaborations, they are situated in the Q1 category

\end{highlights}

\begin{keywords}
Retracted Publication \sep Indian Affiliation  \sep Institutional Collaboration \sep Journal Quartile \sep Retraction Reasons

\end{keywords}

\maketitle

\section{Introduction}

In academia, consistent publication stands as one of the most effective ways to demonstrate one's research capabilities. Successfully publishing research not only garners attention for scholars and their affiliated institutions but also has the potential to attract increased funding and propel individuals' careers within their respective fields. Moreover, many private institutes and universities offer financial incentives for each publication, while a high volume of publications can also enhance annual appraisals. Academic institutions often assess competence based on the number of publications attributed to an individual, with administrators increasingly relying on this metric in recruitment decisions. Scholars who publish infrequently or prioritize activities that do not result in publications may find themselves at a disadvantage when seeking appraisal. These various factors collectively contribute to the phenomenon known as ``Publish or perish,'' a term initially coined by Coolidge in 1932, which has since become an unavoidable reality ~\citep{rawat2014publish}.

Whether driven by the pursuit of monetary rewards or the pressure to publish for academic advancement, these factors have led authors to resort to unethical practices such as fake peer review, data falsification, plagiarism, and false authorship~\citep{racimo2022ethical, receveur2024david}. Such practices have gradually eroded the quality and authenticity of research work. Subsequently, as journals and ethical authors became aware of this trend, they began retracting such unfair and false studies. Currently, the quantity of published papers facing retraction is increasing day by day~\citep{vuong2020characteristics}. While retractions constitute only a tiny fraction of the overall published literature, the alarming rate at which retractions are escalating raises concerns. Retracted papers not only tarnish a researcher's career~\citep{mongeon2016costly, sharma2023ripple} but also undermine the integrity of scholarly literature itself.

In pursuit of academic progress, the majority of Indian higher education institutions expect their faculty, encompassing both academics and researchers, to produce at least one paper in a journal indexed in Scopus or Web of Science. This expectation reflects the widely acknowledged ``publish or perish'' phenomenon~\citep{rawat2014publish, lee2014publish}, where the sheer quantity of publications often outweighs considerations of their quality. Addressing this concern necessitates a shift in academic evaluations to prioritize the substance of publications over the prestige of the journals where they are published~\citep{chaddah2018policy}. Moreover, it underscores that retracted papers stemming from international collaborations are predominantly associated with high-quality journals, while those originating from domestic collaborations tend to be from journals of middling to low quality.

Vuong et al. presented in their study that the frequency of retraction has boomed in the past 20 years and fake peer review accounts for the major reason for retraction~\citep{vuong2020characteristics}. They also stated that the Institute of Electrical and Electronics Engineers, Elsevier, and Springer account for nearly 60\% of all retracted papers globally.  Bar-ilan et al. have explored the temporal characteristics of retracted articles, including time of publication, years to retract, growth of post-retraction citations over time, and social media attention for articles retracted due to ethical misconduct, scientific distortion, and administrative error~\citep{bar2018temporal}.  They found that both citation counts and Mendeley reader counts continue to grow after retraction. Tang et al. found that publications with authors from elite universities are less likely to be retracted, which is particularly true for retractions due to falsification, fabrication, and plagiarism~\citep{tang2020retraction}.

The current review processes organized by journals do not effectively ensure the ethical quality of publications \citep{bornmann2011improving}. One approach used to address this issue is article retraction. If a published article later proves to be unfit for publication due to particularly serious reasons, it may be retracted. However, given the gravity of this action, it must be exercised cautiously to prevent innocent individuals from being adversely affected. There are uncertainties regarding the classification of retraction reasons. Previous studies have presented differing perspectives on the rise in retractions attributed to errors or misconduct~\citep{fanelli2013growing, fang2012misconduct, steen2011retractions_a, campos2019misconduct, fang2012misconduct, helgesson2015plagiarism}.COPE suggests that journal editors should consider retracting an article in any of the following four situations~\citep{kleinert2009cope}: (i) when there is evidence that the findings are unreliable, either due to misconduct or honest error; (ii) when the same findings have already been published without proper acknowledgment (e.g., without obtaining permission or providing a reference); (iii) when the article plagiarizes another published document; or (iv) when the article is based on unethical research~\citep{wager2009retractions}. According to these COPE guidelines, an article could be retracted due to various reasons such as honest errors, misrepresentation, data manipulation, plagiarism (including self-plagiarism), text duplication, or failure to disclose conflicts of interest~\citep{kleinert2009cope, wager2009retractions}.

Considerable literature has examined scientific misconduct within the Indian research landscape. Sharma et al. ~\citep{sharma2023systematic} systematically analyzed reasons for retractions and their citations in the biomedical field for papers affiliated with India. They found that plagiarism, falsification, and fabrication of data are major reasons for retractions in the biomedical field. Elango et al. ~\citep{elango2019analysis} similarly investigated the reasons for retractions among papers affiliated with India, identifying plagiarism as the most common cause. Furthermore, Elango analyzed biomedical literature from Indian authors, revealing that a significant portion of retracted articles were published after 2010, with plagiarism and fake data being common reasons for retractions~\citep{elango2021retracted}. Additionally, more than half of the retracted articles were co-authored within the same institutions, with no repeat offenders identified. The majority of funded research retractions were attributed to fake data, whereas plagiarism was prevalent among non-funded research. Furthermore, Dhingra et al. ~\citep{dhingra2014publication} surveyed the extent of misconduct in publication among biomedical researchers.  Mukhopadhyay et al. analyzed 1376 retracted papers to identify repeated offenders, author distribution, gender distribution, and institute distribution, noting that elite educational and research institutes bear equal responsibility for retractions~\citep{mukhopadhyay2023retractions}. Sabir et al. discussed misconduct in healthcare, highlighting reasons for misconduct and proposing steps to reduce it~\citep{sabir2015scientific}. Khurana et al. have also shown comparative patterns of retractions among USA, China, and India on covid-19 research~\citep{Khurana2024covid19}.

The current literature on the analysis of retractions in India primarily focuses on the biomedical field. For instance, a study conducted by Elango et al.~\citep{elango2019analysis} examined 239 retractions from Scopus within the Indian research landscape spanning from 2005 to 2018. Mukhopadhyay et al. also analyzed 1376 retracted papers listed in Retraction Watch~\citep{mukhopadhyay2023retractions}. In contrast, this study offers a more comprehensive analysis covering more than two decades and a larger dataset of retractions. While previous studies mainly aimed to understand the reasons behind retractions, our study delves into the reasons for retractions about journal quartiles, institutional collaboration, and publishing houses. Additionally, our study provides a detailed examination of retractions in terms of inter-country and intra-country collaborations, in correspondence with retraction reasons and journal quartiles, a perspective not previously explored in either the Indian or global literature on retractions.

\subsection{Research Objectives}
The study aims to analyze Indian retractions with objectives:
\begin{itemize}
\item Examine the year-wise trend of the number of retractions and the duration of retractions (measured in months).
\item Analyze the distribution of retractions based on journal quartile, the reason for retraction,  institution category, and publishing houses.
\item Determine the type of institutions involved in retractions (public, private, or medical), along with the cause of retraction and its impact.
\item Analyze patterns of retractions stemming from intra-country and inter-country institute collaborations, focusing on the journals involved, reasons for retraction, and journal impact.
\end{itemize}

The structure of this study is outlined as follows. Section~\ref{sec:methodology} details the process of data collection, filtration, and categorization of institutions. Section~\ref{sec:results} delves into the findings and discussion, highlighting key aspects of the retraction trajectory, journal quartile, and reasons for retraction. Section~\ref{sec:institute} offers insights into institution coverage and the distribution of author collaborations and institutional partnerships. Additionally, Section~\ref{sec:inter-intra} presents a comparison of inter-country and intra-country institutional collaborations. Section~\ref{sec:country} showcases the collaboration network among countries. Finally, the study concludes with corresponding considerations, limitations, and policy recommendations.

\section{Methodology}
\label{sec:methodology}
\subsection{Data Collection and Filtration}

The primary source of data is the Retraction Watch database obtained via Crossref~\citep{crossref2023} in February 2024. It comprises 51,275 publications, encompassing various details such as title, author names, country affiliations, journal and publisher names, publication subjects, types, retraction specifics including reasons, publication and retraction dates,  paper DOIs, and document type.

To identify publications associated with Indian affiliation, papers were filtered based on the presence of "India" within at least one affiliation, resulting in 3,415 records. To refine the dataset to unique entries, further scrutiny was conducted regarding title duplication, original paper DOI duplication, and retraction types. The dataset included four retraction types: correction, expression of concern, reinstatement, and retraction. The ultimate selection for filtering was ``\textit{Retraction}''.  Consequently, after these refinement steps, the dataset was narrowed down to 3,244 unique retractions.

\subsection{Institute Categorization}

From the initial pool of 3,244 papers, we extracted the department and institution names affiliated with India, culminating in the identification of 5,541 institutions across the 3,244 publications. These institutions underwent further classification according to their nature, including private universities/institutes, private colleges, public universities, institutes of national importance (such as IITs, IIITs, IISc, NITs, etc.) with national research centers (such as CSIR, national labs, etc.), and medical institutes (encompassing hospitals and clinics), adhering to the categorization standards outlined by the University Grants Commission, India (UGC) (\url{https://www.ugc.gov.in/universitydetails/university?type=ddmCMsxJZgXH2S/m0uMOKQ==}). During the data refinement stage, 102 institutes remained unidentifiable due to insufficient affiliation information or abbreviation usage, resulting in a final tally of 5,439 institutes.

\begin{figure}[!h]
    \centering
    \includegraphics[width=0.75\linewidth]{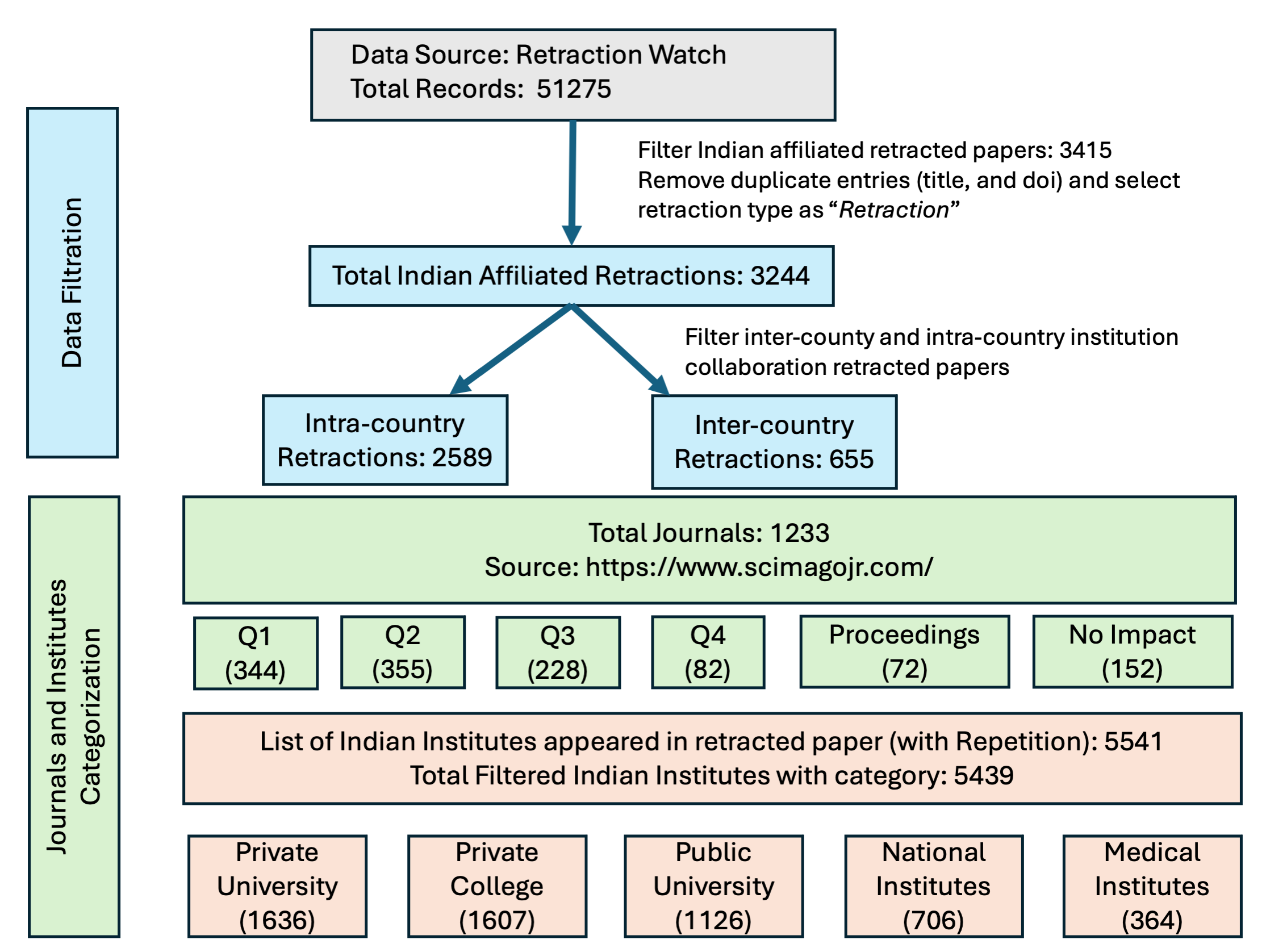}
\caption{Data collection and filtration process of India affiliated retracted publications.}
\label{fig:DataFlow}   
\end{figure}

\section{Results and Discussions}
\label{sec:results}
\subsection{Retraction Trend- Yearly and Monthly}

\begin{figure}[!h]
    \centering
    \includegraphics[width=0.85\linewidth]{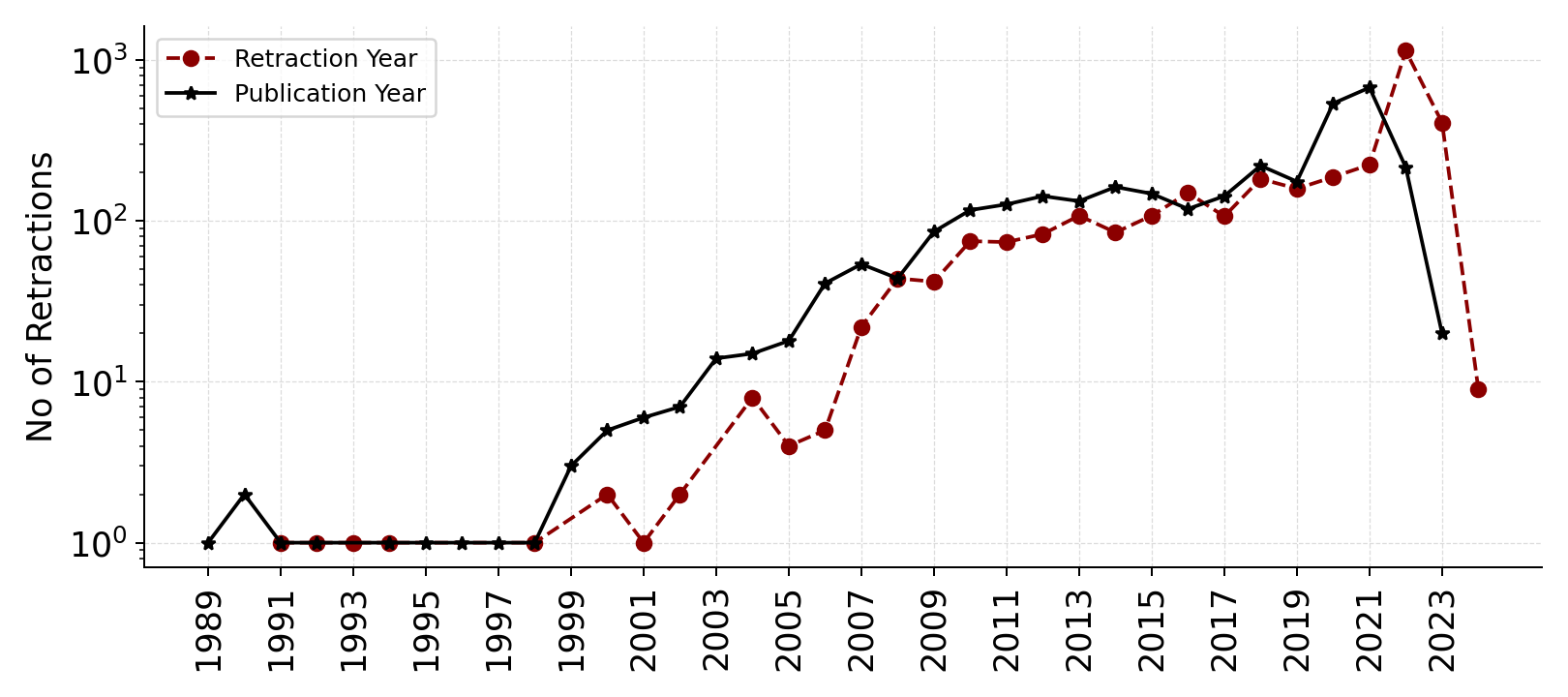}
      \llap{\parbox[b]{5.5in}{(a)\\\rule{0ex}{2.3in}}}
        \includegraphics[width=0.85\linewidth]{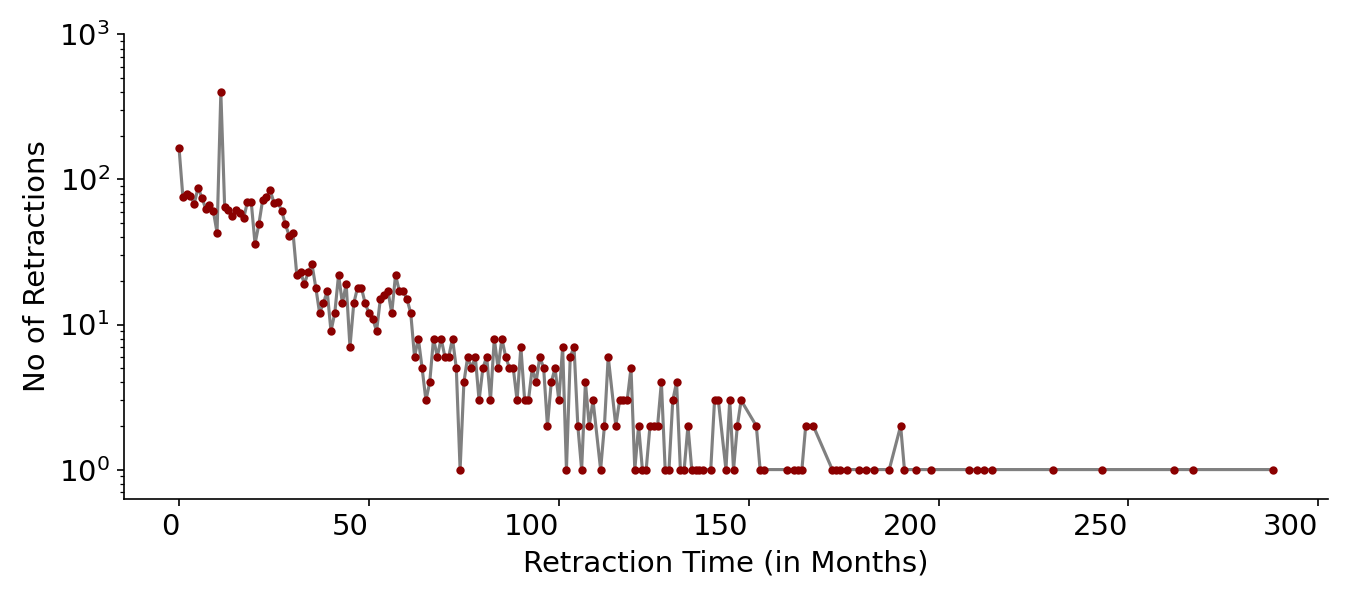}
          \llap{\parbox[b]{5.5in}{(b)\\\rule{0ex}{2.3in}}}
\caption{(a) Publication timeline of retracted papers (solid black) and retraction timeline of published papers (dotted red). (b) Retraction time in months.}
\label{fig:RT}   
\end{figure}

Figure~\ref{fig:RT} (a) illustrates the trajectory of retractions (red dashed line) alongside their corresponding publication dates (black solid line). There was a decline in the number of retractions in 2004, followed by a subsequent increase since 2005. The trend displays a continuous upward pattern, peaking in 2022 before declining. Among papers published during 1990-1999, there were 13 instances, with only 5 retracted before 2000, while the remaining 9 were retracted after 2007. Notably, papers published between 1990 and 1999 tend to be retracted much later, as highlighted by Elango et al. (2019). On average, retractions of papers published during 1990-1999 took approximately 15 years to occur.

In Figure~\ref{fig:RT} (b), a notable proportion of papers are retracted early, indicated by a decreasing trend. The timeline for retractions ranges from 0 to 288 months. Approximately 4.5\% of papers are retracted within a month, 19.4\% in six months, while 40.8\% are retracted within 12 months.

\subsection{Journal Quartile}

Alongside the Impact Factor or Impact Index, journals are categorized into quartiles within each subject category by both Journal Citation Reports (JCR)-Clarivate (\url{https://clarivate.com/}) and Scimago Journal \& Country Rank (SJR) (\url{https://www.scimagojr.com/}).  A quartile represents the ranking of a journal determined by various factors such as impact factor, citation count, and indexing within a specific database~\citep{ahmet2020quartile}. JCR, a paid service, is managed by Web of Science, while SJR, an open-source platform, is maintained by Scopus. These quartiles organize journals from highest to lowest based on their impact index, dividing them into four categories: Q1, Q2, Q3, and Q4. The most esteemed journals within a subject area are typically found in the first quartile, Q1.
\begin{itemize}
\item \textit{Quartile-1 (Q1)} encompasses the top 25\% of journals.
\item \textit{Quartile-2 (Q2) }represents the middle-high position, ranging between 25\% to 50\% of journals.
\item \textit{Quartile-3 (Q3)} signifies the middle-low position, covering journals ranked between 50\% to 75\%.
\item \textit{Quartile-4 (Q4)} indicates the lowest position, encompassing journals ranked between 75\% to 100\%.
\end{itemize}

Additionally,\textit{Proceedings} refers to papers published in conference proceedings following peer review.
\textit{No Impact} denotes papers published in journals not indexed by Scopus or stored in scholarly platforms like arXiv.

\begin{figure}[!h]
    \centering
    \includegraphics[width=0.48\linewidth]{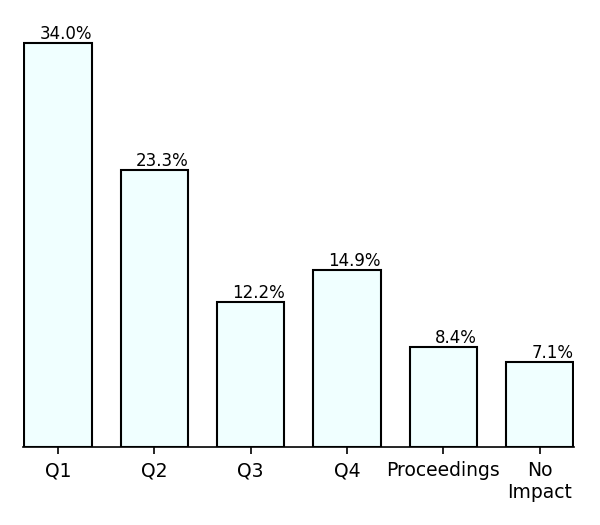}
\caption{Number of retractions as per journal quartile.}
\label{fig:JQ}   
\end{figure}

Figure~\ref{fig:JQ} illustrates the distribution of retractions based on journal quartiles. The highest proportion of retractions originates from Q1 journals (34\%),  followed by Q2 journals (23.3\%). Retractions associated with Q4 journal (14.9\%) are higher than Q3 (12.2\%).
\subsection{Retractions as per Top 10 Publishers}

As illustrated in Table~\ref{table:Publisher_Quartile}, a notable portion of retractions (28.5\%) originates from Springer Publishing House, with 55.1\% retracted from Q1 journals. IOP Publishing ranks second, contributing to 14.03\% of retractions, with 59.8\% originating from Q4 journals and 37.5\% from conference proceedings. The third highest retraction rate is from Elsevier (12.7\%), predominantly from Q1 journals (56.9\%). Hindawi (43.1\%), Taylor and Francis (36.7\%), and Wiley (46.6\%) predominantly experience retractions in Q2 journals. Wolters Kluwer (37.2\%) primarily experiences retractions from Q3 journals, while SAGE Publications (49.5\%) does so from Q4 journals. All retractions from the PLoS publishing house are associated with Q1 impact. Additionally, Hindawi (10.3\%) and Wolters Kluwer (14.7\%) experience significant instances of retractions with no impact. Retractions from IEEE predominantly involve proceedings (55.5\%)~\citep{vuong2020characteristics},  followed by IOP Publishing (37.5\%). This indicates that retractions from Q1 journals primarily originated from Springer and Elsevier, while those from Q2 journals came mainly from Hindawi, Taylor and Francis, and Wiley. Retractions from Q3 journals were associated with Wolters Kluwer, and those from Q4 journals were predominantly from SAGE Publications and IOP Publishing.

\begin{table}[]
\caption{Number of retractions as per journal quartile for top 10 publishers.}
\begin{tabular}{|l|ccc|ccccc|}
\hline
\multirow{2}{*}{\textbf{Publisher}} &
  \multicolumn{3}{c|}{\textbf{No of Retractions}} &
  \multicolumn{5}{c|}{\textbf{Journal Quartile (in\%)}} \\ \cline{2-9} 
 &
  \multicolumn{1}{c|}{\textbf{Count}} &
  \multicolumn{1}{c|}{\textbf{In\%}} &
  \textbf{Q1} &
  \multicolumn{1}{c|}{\textbf{Q2}} &
  \multicolumn{1}{c|}{\textbf{Q3}} &
  \multicolumn{1}{c|}{\textbf{Q4}} &
  \multicolumn{1}{c|}{\textbf{Proceedings}} &
  \textbf{No Impact} \\ \hline
Springer &
  \multicolumn{1}{c|}{927} &
  \multicolumn{1}{c|}{28.58} &
  55.12 &
  \multicolumn{1}{c|}{26.43} &
  \multicolumn{1}{c|}{8.63} &
  \multicolumn{1}{c|}{2.59} &
  \multicolumn{1}{c|}{2.59} &
  4.64 \\ \hline
IOP Publishing &
  \multicolumn{1}{c|}{455} &
  \multicolumn{1}{c|}{14.03} &
  - &
  \multicolumn{1}{c|}{2.64} &
  \multicolumn{1}{c|}{-} &
  \multicolumn{1}{c|}{59.78} &
  \multicolumn{1}{c|}{37.58} &
  - \\ \hline
Elsevier &
  \multicolumn{1}{c|}{413} &
  \multicolumn{1}{c|}{12.73} &
  56.9 &
  \multicolumn{1}{c|}{24.94} &
  \multicolumn{1}{c|}{10.65} &
  \multicolumn{1}{c|}{1.94} &
  \multicolumn{1}{c|}{0.97} &
  4.6 \\ \hline
Hindawi &
  \multicolumn{1}{c|}{204} &
  \multicolumn{1}{c|}{6.29} &
  23.04 &
  \multicolumn{1}{c|}{43.14} &
  \multicolumn{1}{c|}{20.1} &
  \multicolumn{1}{c|}{3.43} &
  \multicolumn{1}{c|}{-} &
  10.29 \\ \hline
Wolters Kluwer &
  \multicolumn{1}{c|}{156} &
  \multicolumn{1}{c|}{4.81} &
  7.05 &
  \multicolumn{1}{c|}{28.85} &
  \multicolumn{1}{c|}{37.18} &
  \multicolumn{1}{c|}{12.18} &
  \multicolumn{1}{c|}{-} &
  14.74 \\ \hline
Taylor and Francis &
  \multicolumn{1}{c|}{136} &
  \multicolumn{1}{c|}{4.19} &
  26.47 &
  \multicolumn{1}{c|}{36.76} &
  \multicolumn{1}{c|}{22.06} &
  \multicolumn{1}{c|}{12.5} &
  \multicolumn{1}{c|}{-} &
  2.21 \\ \hline
IEEE &
  \multicolumn{1}{c|}{126} &
  \multicolumn{1}{c|}{3.88} &
  11.9 &
  \multicolumn{1}{c|}{10.32} &
  \multicolumn{1}{c|}{15.87} &
  \multicolumn{1}{c|}{6.35} &
  \multicolumn{1}{c|}{55.56} &
  - \\ \hline
Wiley &
  \multicolumn{1}{c|}{120} &
  \multicolumn{1}{c|}{3.7} &
  27.5 &
  \multicolumn{1}{c|}{46.67} &
  \multicolumn{1}{c|}{19.17} &
  \multicolumn{1}{c|}{-} &
  \multicolumn{1}{c|}{-} &
  6.67 \\ \hline
SAGE Publications &
  \multicolumn{1}{c|}{99} &
  \multicolumn{1}{c|}{3.05} &
  7.07 &
  \multicolumn{1}{c|}{33.33} &
  \multicolumn{1}{c|}{8.08} &
  \multicolumn{1}{c|}{49.49} &
  \multicolumn{1}{c|}{1.01} &
  1.01 \\ \hline
PLoS &
  \multicolumn{1}{c|}{82} &
  \multicolumn{1}{c|}{2.53} &
  100 &
  \multicolumn{1}{c|}{-} &
  \multicolumn{1}{c|}{-} &
  \multicolumn{1}{c|}{-} &
  \multicolumn{1}{c|}{-} &
  - \\ \hline
\end{tabular}
\label{table:Publisher_Quartile}
\end{table}

\subsection{Retraction Reasons}

Studies can face retraction for a variety of reasons, broadly categorized into honest error, intentional or unintentional misconduct, and other factors~\citep{xu2022cross, sharma2023systematic}. Honest errors typically involve mistakes in sample or data collection, irreproducibility of results, disputes over authorship attribution, redundant publication, and similar issues. On the other hand, misconduct encompasses actions such as plagiarism, undisclosed conflicts of interest, data fabrication or manipulation, ethical concerns, duplicate submissions, and related misconduct. The third category, labeled as ``others'', encompasses retractions where the rationale is either not explicitly stated in the retraction notice or remains unclear to the journal editors. For example, authors may retract a paper that has been simultaneously submitted to and accepted by another journal, which constitutes misconduct for failing to inform the journal about the duplicate submission. However, if authors do not specify the reason for retraction, it would be categorized under ``other''. Based on the aforementioned categories, the provided data is labeled for the following retraction reasons.
\begin{itemize}
\item \textit{Fake Peer Review}: This involves recommending peer reviewers to journals upon paper submission, providing falsified contact information, and subsequently submitting fraudulent, excessively favorable peer review reports, typically resulting in fast publication~\citep{ferguson2014peer, watch2017springer-bmc}.

\item \textit{Data Integrity}: This encompasses errors in data, fabricated results, data manipulation, and content generated randomly, among other issues~\citep{chambers2019plagiarism}.

\item \textit{Plagiarism}: Plagiarism occurs when someone uses another person's intellectual work, such as texts, ideas, or results, without proper citation, implying it to be their own~\citep{helgesson2015plagiarism}.

\item \textit{Duplication of Article}: This refers to the publication of the same paper in multiple journals~\citep{alfonso2005duplicate}.

\item \textit{Authors Related}: This arises when individuals are listed as authors on a published manuscript without their participation or consent~\citep{hosseini2020review}.

\item \textit{Ethical Concerns}: Ethical concerns typically arise from a failure to obtain ethical approval before conducting a study involving animal and/or human subjects~\citep{elia2014fate}.

\item \textit{Error by Journal}: This occurs when a journal mistakenly publishes the same paper again.

\item \textit{Multiple}: This category encompasses various reasons for retraction, such as authorship disputes, conflicts of interest, data manipulation, and copyright issues, among others.

\item \textit{Unknown}: This category includes cases where the specific reason for retraction is not known.
\end{itemize}

In Table~\ref{table:Ret_reason_Quartile}, it becomes apparent that the primary reason for the majority of retractions is fake peer review, constituting 33\% of the total retractions. Papers retracted from the Q1 category are mainly due to fake peer reviews (41.9\%). Retractions from Q2 journals are nearly equally distributed among fake peer review (22\%), data integrity (20.5\%), and multiple reasons (21.9\%). For Q3 retractions, the majority are due to plagiarism (21.7\%) and data integrity issues (20.7\%). In Q4, the majority are due to fake peer reviews (58.8\%). Additionally, conference proceedings are primarily retracted due to fake peer reviews (41.9\%).
The second-highest cause of retraction in Q1 journals is attributed to data integrity issues (17.2\%), with 19.6\% of retractions originating from Q1 journals and 20.4\% from Q2 journals. Plagiarism accounts for 14.8\% of total retractions. Notably, 32\% retraction is due to plagiarism and is from journals not listed in the Scopus, thus having no impact. Furthermore, 18.2\% of retractions stem from multiple reasons, with nearly equal proportions from Q1, Q2, Q3 journals, and conference proceedings. Additionally, 12.9\% of retractions from Q3 journals are attributed to article duplication. Combined, fake peer review, data integrity, and plagiarism account for approximately 65\% of retractions.

\begin{table}[!h]
\caption{Number of retraction as per journal quartile and their corresponding reason of retraction.}
\begin{tabular}{|l|ccc|ccccc|}
\hline
\multirow{2}{*}{\textbf{Retraction Reasons}} &
  \multicolumn{3}{c|}{\textbf{No of Retractions}} &
  \multicolumn{5}{c|}{\textbf{Journal Quartile (in \%)}} \\ \cline{2-9} 
 &
  \multicolumn{1}{c|}{\textbf{Count}} &
  \multicolumn{1}{c|}{\textbf{In \%}} &
  \textbf{Q1} &
  \multicolumn{1}{c|}{\textbf{Q2}} &
  \multicolumn{1}{c|}{\textbf{Q3}} &
  \multicolumn{1}{c|}{\textbf{Q4}} &
  \multicolumn{1}{c|}{\textbf{Proceedings}} &
  \textbf{No Impact} \\ \hline
Fake Peer Review &
  \multicolumn{1}{c|}{1071} &
  \multicolumn{1}{c|}{33.01} &
  41.94 &
  \multicolumn{1}{c|}{22.06} &
  \multicolumn{1}{c|}{10.63} &
  \multicolumn{1}{c|}{58.8} &
  \multicolumn{1}{c|}{41.97} &
  - \\ \hline
Multiple &
  \multicolumn{1}{c|}{589} &
  \multicolumn{1}{c|}{18.16} &
  18.84 &
  \multicolumn{1}{c|}{21.93} &
  \multicolumn{1}{c|}{18.48} &
  \multicolumn{1}{c|}{9.94} &
  \multicolumn{1}{c|}{19.34} &
  17.75 \\ \hline
Data Integrity &
  \multicolumn{1}{c|}{559} &
  \multicolumn{1}{c|}{17.23} &
  19.66 &
  \multicolumn{1}{c|}{20.48} &
  \multicolumn{1}{c|}{20.76} &
  \multicolumn{1}{c|}{12.22} &
  \multicolumn{1}{c|}{6.2} &
  12.55 \\ \hline
Plagiarism &
  \multicolumn{1}{c|}{481} &
  \multicolumn{1}{c|}{14.83} &
  10.69 &
  \multicolumn{1}{c|}{16.91} &
  \multicolumn{1}{c|}{21.77} &
  \multicolumn{1}{c|}{6.63} &
  \multicolumn{1}{c|}{15.69} &
  32.03 \\ \hline
Unknown &
  \multicolumn{1}{c|}{188} &
  \multicolumn{1}{c|}{5.8} &
  2.26 &
  \multicolumn{1}{c|}{4.36} &
  \multicolumn{1}{c|}{7.34} &
  \multicolumn{1}{c|}{5.8} &
  \multicolumn{1}{c|}{7.66} &
  22.51 \\ \hline
Duplication of Article &
  \multicolumn{1}{c|}{180} &
  \multicolumn{1}{c|}{5.55} &
  3.08 &
  \multicolumn{1}{c|}{7.66} &
  \multicolumn{1}{c|}{12.91} &
  \multicolumn{1}{c|}{3.31} &
  \multicolumn{1}{c|}{1.09} &
  7.79 \\ \hline
Authors Related &
  \multicolumn{1}{c|}{86} &
  \multicolumn{1}{c|}{2.65} &
  1.99 &
  \multicolumn{1}{c|}{3.3} &
  \multicolumn{1}{c|}{3.8} &
  \multicolumn{1}{c|}{1.66} &
  \multicolumn{1}{c|}{1.46} &
  5.19 \\ \hline
Error by Journal &
  \multicolumn{1}{c|}{49} &
  \multicolumn{1}{c|}{1.51} &
  1.09 &
  \multicolumn{1}{c|}{1.98} &
  \multicolumn{1}{c|}{2.53} &
  \multicolumn{1}{c|}{1.45} &
  \multicolumn{1}{c|}{0.73} &
  1.3 \\ \hline
Ethical Concerns &
  \multicolumn{1}{c|}{41} &
  \multicolumn{1}{c|}{1.26} &
  0.45 &
  \multicolumn{1}{c|}{1.32} &
  \multicolumn{1}{c|}{1.77} &
  \multicolumn{1}{c|}{0.21} &
  \multicolumn{1}{c|}{5.84} &
  0.87 \\ \hline
\textbf{Grand Total} &
  \multicolumn{1}{c|}{\textbf{3244}} &
  \multicolumn{1}{c|}{\textbf{100}} &
  \textbf{1104} &
  \multicolumn{1}{c|}{\textbf{757}} &
  \multicolumn{1}{c|}{\textbf{395}} &
  \multicolumn{1}{c|}{\textbf{483}} &
  \multicolumn{1}{c|}{\textbf{274}} &
  \textbf{231} \\ \hline
\end{tabular}
\label{table:Ret_reason_Quartile}
\end{table}

\subsection{Average Retraction Time}

As indicated in Table~\ref{table:avg_retractTime}, the shortest average retraction period is observed for papers published in conference proceedings (12.17 months), while retractions from Q2 publications took the longest time (32.8 months), followed by Q1 (30.7 months). Fake peer review, being the highest mode of retraction, took an average of 21.5 months, while data integrity ranks as the second most common reason for retraction, with the highest average retraction period of 38.8 months. Similarly, duplication of articles, despite being less frequent in retractions, also took a long time at 33.4 months. Moreover, most retractions are of the article type, with an average retraction time of 31.6 months. Conversely, conference proceedings have the shortest retraction time at 10.9 months. Reviews, articles, and clinical studies took nearly equal durations for retraction.
\begin{table}[!h]
\caption{ Average time of retraction by journal quartile, retraction reasons, and document type..}
\begin{tabular}{|l|c|l|l|c|l|l|c|}
\hline
\textbf{\begin{tabular}[c]{@{}l@{}}Journal \\ Quartile\end{tabular}} &
  \textbf{\begin{tabular}[c]{@{}c@{}}Avg Retraction \\ Time\\  (in months)\end{tabular}} &
  \multirow{10}{*}{\textbf{}} &
  \textbf{\begin{tabular}[c]{@{}l@{}}Retraction \\ Reasons\end{tabular}} &
  \textbf{\begin{tabular}[c]{@{}c@{}}Avg Retraction \\ Time \\ (in months)\end{tabular}} &
  \multirow{10}{*}{\textbf{}} &
  \textbf{\begin{tabular}[c]{@{}l@{}}Document \\ Type\end{tabular}} &
  \textbf{\begin{tabular}[c]{@{}c@{}}Avg Retraction \\ Time \\ (in months)\end{tabular}} \\ \cline{1-2} \cline{4-5} \cline{7-8} 
Q1          & 30.71      &  & Data Integrity         & 38.87 &  & Review         & 34.14      \\ \cline{1-2} \cline{4-5} \cline{7-8} 
Q2          & 32.85      &  & Duplication of Article & 33.47 &  & Article        & 31.68      \\ \cline{1-2} \cline{4-5} \cline{7-8} 
Q3          & 25.47      &  & Multiple               & 32.64 &  & Clinical Study & 30.54      \\ \cline{1-2} \cline{4-5} \cline{7-8} 
Q4          & 23.99      &  & Plagiarism             & 27.93 &  & Case Report    & 23.66      \\ \cline{1-2} \cline{4-5} \cline{7-8} 
Proceedings & 12.17     &  & Error by Journal       & 26.67 &  & Book Chapter   & 22.83      \\ \cline{1-2} \cline{4-5} \cline{7-8} 
No Impact   & 24.55       &  & Fake Peer Review       & 21.52 &  & Others         & 16.93      \\ \cline{1-2} \cline{4-5} \cline{7-8} 
\textbf{-}  & \textbf{-} &  & Authors Related        & 21.14 &  & Commentary     & 13.33      \\ \cline{1-2} \cline{4-5} \cline{7-8} 
-           & -          &  & Ethical Concerns       & 14.65 &  & Conference     & 10.99      \\ \cline{1-2} \cline{4-5} \cline{7-8} 
-           & -          &  & Unknown                & 11.85 &  & \textbf{-}     & \textbf{-} \\ \hline
\end{tabular}
\label{table:avg_retractTime}
\end{table}
\section{Institution Categorization and Collaboration}
\label{sec:institute}

\begin{figure}[!h]
    \centering
    \includegraphics[width=0.45\linewidth]{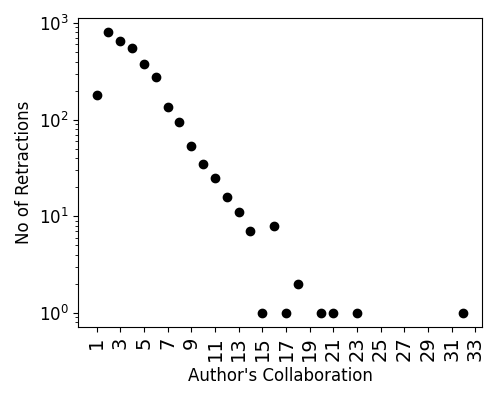}
          \llap{\parbox[b]{0.5in}{(a)\\\rule{0ex}{2.0in}}}
        \includegraphics[width=0.45\linewidth]{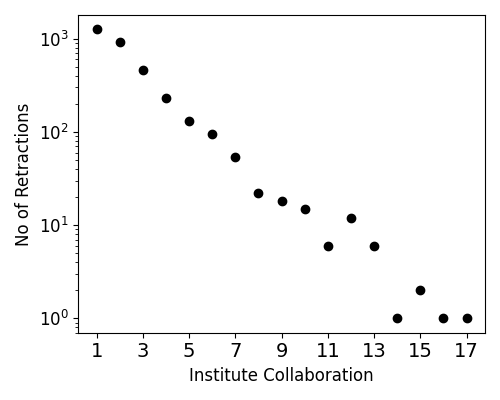}
           \llap{\parbox[b]{0.5in}{(b)\\\rule{0ex}{2.0in}}}
\caption{(a) Author's collaboration count. (b) Institute collaboration count.}
\label{fig:colab}   
\end{figure}

Figure~\ref{fig:colab} shows that approximately 60\% of retractions stem from private colleges, universities, and institutes, with 33.7\% originating from public universities, institutes, and research centers. A smaller proportion, about 6.7\%, comes from medical institutes, hospitals, and clinics. 
When categorizing institutes, ``National Institutes'' includes all establishments of national significance such as IIT, IIIT, NIT, ISI, ISC, etc., along with research centers like CSIR and national laboratories. ``Medical institutes'' encompass all nationally important medical establishments such as All India Institute of Medical Sciences (AIIMS), Post Graduate Institute of Medical Education \& Research (PGIMER), etc., as well as hospitals like Apollo Hospitals and Sir Ganga Ram Hospital.
\begin{figure}[!h]
    \centering
    \includegraphics[width=0.48\linewidth]{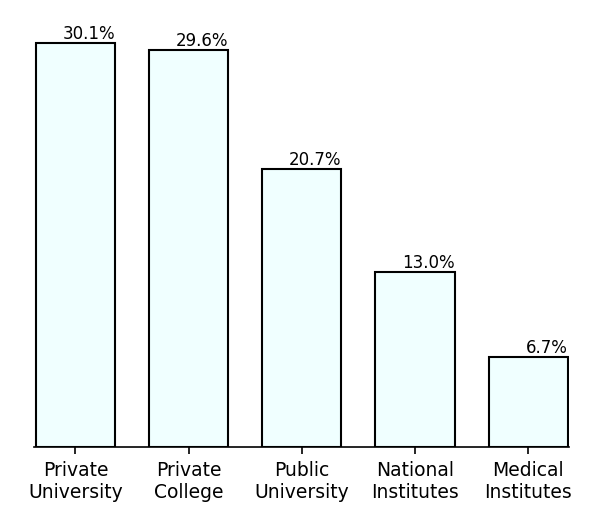}
\caption{Number of retractions as per institute category.}
\label{fig:Fig3}   
\end{figure}

The majority of retractions occur within smaller team sizes, with only 5.6\% of retractions attributed to single authors. The highest proportion of retractions is observed in teams consisting of two authors (25.1\%), followed by three authors (19.9\%) and four authors (17.1\%). Moreover, there are individual papers with author sizes of 15, 17, 20, 21, 23, and 32. The trend depicted in Figure~\ref{fig:colab} (a) suggests that as the team size increases, the number of retractions decreases~\cite{sharma2021team}. Approximately 50.6\% of retractions occur when up to three authors collaborate, and this increases to 87.9\% when up to six authors collaborate.

Likewise, Figure~\ref{fig:colab} (b) depicts the pattern of institute collaboration. Fewer retracted papers are affiliated with multiple institutions. The highest number of retractions (38.7\%) occurs when authors are from the same institute, followed by collaborations between two institutes (, and three institutes (14.4\%). Retractions involving collaborations among multiple institutes are minimal. Approximately 67.4\% of retractions occur when up to two institutes collaborate, increasing to 92.8\% when up to five institutes collaborate.  In single institute retraction, majority of the retraction is from private colleges (24.5\%) followed by private universities (22\%). The same pattern is observe in 2 institute collaboration and so on.

\subsection{Retraction Reasons Corresponding to Institution Category}

As indicated in Table~\ref{table:Ret_reason_Institute}, among 1636 instances of appearances in private institutes and universities, 43.2\% of retractions are attributed to fake peer review, with 22.3\% attributed to multiple reasons. Similarly, out of 1607 instances of appearances in private colleges in all retracted publications, 52.7\% of retractions are due to fake peer reviews, with 15.3\% due to multiple reasons. Among public universities, totaling 1126 appearances, the primary cause of retractions is data integrity issues (24.2\%), followed by multiple reasons at 25.4\%. Similarly, in national institutes and research centers comprising 706 appearances, 35.4\% of retractions are linked to data integrity, followed by 20.6\% due to multiple reasons. Medical institutes experience 28.8\% retractions due to data integrity and 19.7\% due to plagiarism. This suggests that fake reviews are prevalent in private institutions, while data integrity issues are common in public institutions.
 
\begin{table}[!h]
\caption{Retraction reasons corresponding to affiliation category (in \%). The number of institutes in each category are with repetition as same institute may appear in multiple publications.}
\begin{tabular}{|l|c|c|c|c|c|}
\hline
\textbf{Retraction Reasons} &
  \textbf{\begin{tabular}[c]{@{}c@{}}Private \\ University\end{tabular}} &
  \textbf{\begin{tabular}[c]{@{}c@{}}Private \\ College\end{tabular}} &
  \textbf{\begin{tabular}[c]{@{}c@{}}Public \\ University\end{tabular}} &
  \textbf{\begin{tabular}[c]{@{}c@{}}National \\ Institutes\end{tabular}} &
  \textbf{\begin{tabular}[c]{@{}c@{}}Medical \\ Institutes\end{tabular}} \\ \hline
Fake Peer Review          & 43.28         & 52.77         & 19.8          & 7.93         & 2.75         \\ \hline
Multiple                  & 22.31         & 15.37         & 25.49         & 20.68        & 15.66        \\ \hline
Data Integrity            & 14.24         & 6.97          & 24.25         & 35.41        & 28.85        \\ \hline
Plagiarism                & 11.74         & 12.69         & 15.45         & 13.17        & 19.78        \\ \hline
Duplication of Article    & 3.48          & 5.1           & 5.15          & 8.22         & 12.09        \\ \hline
Unknown                   & 1.71          & 3.98          & 4.62          & 7.37         & 8.79         \\ \hline
Authors Related           & 1.65          & 1.56          & 2.58          & 3.97         & 6.04         \\ \hline
Error by Journal          & 0.67          & 0.93          & 1.69          & 1.98         & 3.3          \\ \hline
Ethical Concerns          & 0.92          & 0.62          & 0.98          & 1.27         & 2.75         \\ \hline
\textbf{Total Institutes} & \textbf{1636} & \textbf{1607} & \textbf{1126} & \textbf{706} & \textbf{364} \\ \hline
\end{tabular}
\label{table:Ret_reason_Institute}
\end{table}

\section{Intra-country Vs. Inter-country Retractions }
\label{sec:inter-intra}

\begin{figure}[!h]
    \centering
    \includegraphics[width=0.85\linewidth]{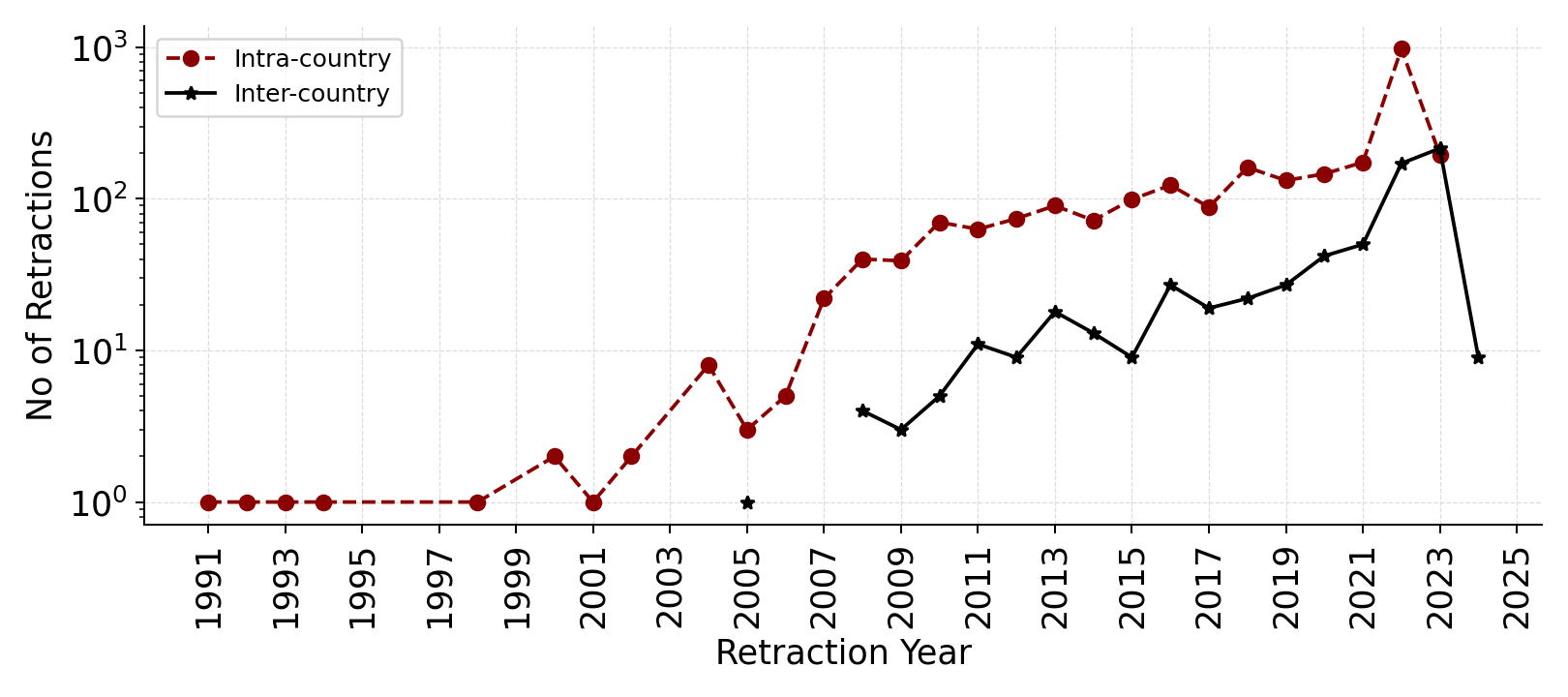}
\caption{Retraction trend of papers with intra-country institutions collaboration (red dotted line) and inter-country institutions collaboration (black solid line).}
\label{fig:Inter-Intra-RT}   
\end{figure}

Intra-country collaborations encompass retractions wherein all authors are affiliated with Indian institutes or universities. Approximately 80\% of the retracted papers exclusively feature Indian affiliations and collaborations. As depicted in Figure~\ref{fig:Inter-Intra-RT}, retractions within the country display an upward trajectory from 2001 to 2004, followed by a decline in 2005. Subsequently, there is an upward trend post-2005, indicating an increase in retractions involving authors collaborating within India. Within the last five years (2019-2023), 62.6\% of retractions occurred.

Conversely, in inter-country collaborations, there is an ascending trend from 2017 to 2023, with fluctuations observed before 2017. Several retractions corresponding to 2024 involve international collaboration. Notably, no retractions were reported from 2000 to 2004 and from 2006 to 2007. Within the last five years (2019-2023), 77\% of retractions occurred.

\subsection{Variation in Journal Quartile}

As depicted in Table~\ref{table:inter-intra-quartile}, retractions classified under Q1 are more prevalent in inter-country collaborations (48.2\%) compared to intra-country collaborations (30.4\%). Similarly, Q2 retractions are higher in inter-country collaborations (29\%) than in intra-country collaborations (21.9\%). Conversely, Q3 and Q4 retractions show higher frequencies in intra-country collaborations (12.3\%, 17.6\%) compared to inter-country collaborations (5.2\%, 11.7\%). In contrast to inter-country collaborations, intra-country collaborations exhibit a higher proportion of retractions in proceedings (10\% vs. 3.8\%). Overall, intra-country collaborations account for 52.3\%—approximately half of the total retractions—from the top two journal quartiles Q1 and Q2, while in inter-country collaborations, 77.2\%—roughly one-third of retractions—originate from the top two quartiles.

\begin{table}[!h]
\caption{Intra-country Vs. inter-country retractions as per journal quartile.}
\begin{tabular}{|l|cc|cc|}
\hline
\multirow{2}{*}{\textbf{\begin{tabular}[c]{@{}l@{}}Journal   \\ Quartile\end{tabular}}} &
  \multicolumn{2}{c|}{\textbf{\begin{tabular}[c]{@{}c@{}}Inter-country   \\ Institute Retractions\end{tabular}}} &
  \multicolumn{2}{c|}{\textbf{\begin{tabular}[c]{@{}c@{}}Intra-country  \\  Institute Retractions\end{tabular}}} \\ \cline{2-5} 
            & \multicolumn{1}{c|}{\textbf{Count}} & \textbf{In   \%} & \multicolumn{1}{c|}{\textbf{Count}} & \textbf{In   \%} \\ \hline
Q1          & \multicolumn{1}{c|}{316}            & 48.24            & \multicolumn{1}{c|}{788}            & 30.44            \\ \hline
Q2          & \multicolumn{1}{c|}{190}            & 29.01            & \multicolumn{1}{c|}{567}            & 21.9             \\ \hline
Q3          & \multicolumn{1}{c|}{34}             & 5.19             & \multicolumn{1}{c|}{318}            & 12.28            \\ \hline
Q4          & \multicolumn{1}{c|}{77}             & 11.76            & \multicolumn{1}{c|}{458}            & 17.69            \\ \hline
Proceedings & \multicolumn{1}{c|}{25}             & 3.82             & \multicolumn{1}{c|}{261}            & 10.08            \\ \hline
No Impact   & \multicolumn{1}{c|}{13}             & 1.98             & \multicolumn{1}{c|}{197}            & 7.61             \\ \hline
\end{tabular}
\label{table:inter-intra-quartile}
\end{table}
\subsection{Variation in Retraction Reasons}

Table~\ref{table:inter-intra-reason} reveals that the primary reason for retraction in intra-country scenarios is fake peer review (35.3\%), whereas, in inter-country cases, retractions occur due to a multitude of reasons (32.2\%). plagiarism (16.2\%) and data integrity issues (16.2\%) constitute the secondary cause of retractions in intra-country situations, while in inter-country collaborations, fake peer review (23.8\%) emerges as the secondary cause, followed by data integrity concerns (21.5\%) as the third most common reason. Issues about authors are more prevalent in inter-country collaborations compared to intra-country ones, whereas article duplication is more prevalent in intra-country collaborations than in inter-country ones.

\begin{table}[!h]
\caption{Intra-country Vs. inter-country retractions as per retraction reasons.}
\begin{tabular}{|l|cc|cc|}
\hline
\multirow{2}{*}{\textbf{Retraction Reasons}} &
  \multicolumn{2}{c|}{\textbf{\begin{tabular}[c]{@{}c@{}}Inter-country   \\ Institute Retractions\end{tabular}}} &
  \multicolumn{2}{c|}{\textbf{\begin{tabular}[c]{@{}c@{}}Intra-country   \\ Institute Retractions\end{tabular}}} \\ \cline{2-5} 
                       & \multicolumn{1}{c|}{\textbf{Count}} & \textbf{In \%} & \multicolumn{1}{c|}{\textbf{Count}} & \textbf{In \%} \\ \hline
Fake Peer Review       & \multicolumn{1}{c|}{156}            & 23.82          & \multicolumn{1}{c|}{915}            & 35.34          \\ \hline
Plagiarism             & \multicolumn{1}{c|}{62}             & 9.47           & \multicolumn{1}{c|}{419}            & 16.18          \\ \hline
Data Integrity         & \multicolumn{1}{c|}{141}            & 21.53          & \multicolumn{1}{c|}{418}            & 16.15          \\ \hline
Multiple               & \multicolumn{1}{c|}{211}            & 32.21          & \multicolumn{1}{c|}{378}            & 14.6           \\ \hline
Unknown                & \multicolumn{1}{c|}{20}             & 3.05           & \multicolumn{1}{c|}{168}            & 6.49           \\ \hline
Duplication of Article & \multicolumn{1}{c|}{24}             & 3.66           & \multicolumn{1}{c|}{156}            & 6.03           \\ \hline
Authors Related        & \multicolumn{1}{c|}{27}             & 4.12           & \multicolumn{1}{c|}{59}             & 2.28           \\ \hline
Error by Journal       & \multicolumn{1}{c|}{6}              & 0.92           & \multicolumn{1}{c|}{43}             & 1.66           \\ \hline
Ethical Concerns       & \multicolumn{1}{c|}{8}              & 1.22           & \multicolumn{1}{c|}{33}             & 1.27           \\ \hline
\end{tabular}
\label{table:inter-intra-reason}
\end{table}

\subsection{Variation in Document Type}
\label{sec:DocType}

In Table~\ref{table:intra_docType}, within intra-country collaborative retractions, articles account for 65.3\% of retractions. The second highest source of retractions is conference proceedings (20.9\%), of which 51.5\% are from quartile Q4 and 36.4\% are listed as proceedings with no impact. Reviews contribute 4.4\% overall, predominantly published in Q2 and Q3. Clinical studies and case reports fall predominantly into the Q3 and Q4 categories. Conversely, in inter-country collaborative retracted papers (refer to Table~\ref{table:inter_docType}), articles constitute 87\% of retractions, with 52.1\% from Q1 journals and 29.8\% from Q2. The second highest retraction source is reviews (5.8\%), with the majority occurring in Q2 and Q3 journals. The majority of clinical studies are published in Q1 journals.
\begin{table}[!h]
\caption{Intra-country: Number of retractions with respect to document type and journal quartile.}
\begin{tabular}{|l|c|c|c|c|c|c|c|c|}
\hline
\textbf{Document Type}     & \textbf{Count} & \textbf{In \%} & \textbf{Q1}  & \textbf{Q2}  & \textbf{Q3}  & \textbf{Q4}  & \textbf{Proceedings} & \textbf{No Impact} \\ \hline
Article        & 1691 & 65.31 & 42.81 & 27.91 & 12.48 & 8.4   & 0.89  & 7.51  \\ \hline
Conference     & 543  & 20.97 & 2.58  & 4.05  & 4.05  & 51.57 & 36.46 & 1.29  \\ \hline
Review         & 114  & 4.4   & 18.42 & 25.44 & 24.56 & 8.77  & 1.75  & 21.05 \\ \hline
Clinical Study & 93   & 3.59  & 18.28 & 25.81 & 31.18 & 12.9  & -     & 11.83 \\ \hline
Case Report    & 65   & 2.51  & 7.69  & 24.62 & 35.38 & 18.46 & -     & 13.85 \\ \hline
Commentary     & 53   & 2.05  & 3.77  & 3.77  & 5.66  & 1.89  & 83.02 & 1.89  \\ \hline
Book Chapter   & 19   & 0.73  & -     & -     & -     & -     & 10.53 & 89.47 \\ \hline
Others         & 11   & 0.42  & 45.45 & 18.18 & 18.18 & 9.09  & -     & 9.09  \\ \hline
\textbf{Total Retractions} & \textbf{2589}  & \textbf{100}   & \textbf{788} & \textbf{567} & \textbf{318} & \textbf{458} & \textbf{261}         & \textbf{197}       \\ \hline
\end{tabular}
\label{table:intra_docType}
\end{table}
\begin{table}[!h]
\caption{Inter-country: Number of retractions with respect to document type and journal quartile.}
\begin{tabular}{|l|c|c|c|c|c|c|c|c|}
\hline
\textbf{Document Type}     & \textbf{Count} & \textbf{In \%} & \textbf{Q1}  & \textbf{Q2}  & \textbf{Q3} & \textbf{Q4} & \textbf{Proceedings} & \textbf{No Impact} \\ \hline
Article        & 570 & 87.02 & 52.11 & 29.82 & 11.05 & 2.63  & -     & 4.39 \\ \hline
Review         & 38  & 5.8   & 15.79 & 44.74 & 31.58 & 5.26  & -     & 2.63 \\ \hline
Conference     & 15  & 2.29  & -     & -     & -     & 26.67 & 66.67 & 6.67 \\ \hline
Clinical Study & 13  & 1.98  & 46.15 & 0.53  & 1.3   & 12    & -     & 5.88 \\ \hline
Case Report    & 6   & 0.92  & 1.27  & -     & 0.32  &       & -     & 0.32 \\ \hline
Commentary     & 5   & 0.76  & 20    & -     & -     & 20    & 60    & -    \\ \hline
Others         & 4   & 0.61  & 50    & 50    & -     & -     & -     & -    \\ \hline
Book Chapter   & 4   & 0.61  & -     & -     & -     & -     & -     & 100  \\ \hline
\textbf{Total Retractions} & \textbf{655}   & \textbf{100}   & \textbf{316} & \textbf{190} & \textbf{77} & \textbf{25} & \textbf{13}          & \textbf{34}        \\ \hline
\end{tabular}
\label{table:inter_docType}
\end{table}

\subsection{Variation in Publishing Journals}

Table~\ref{table:intra_Journal} displays the top 10 journals from which intra-collaborative papers were initially published and later retracted. The majority of the published journals belonged to the Q2 and Q4 categories across various publishing houses. However, the highest number of retractions (12.9\%) originated from Q1 journals published by Springer. The second-highest source of retractions was conference proceedings published by IOP Publishing and listed as Q4. As mentioned earlier in Section~\ref{sec:DocType}, intra-country retractions are predominantly articles and conference proceedings, with a majority originating from \textit{Journal of Ambient Intelligence and Humanized Computing} and \textit{Journal of Physics: Conference Series}.

Similarly, Table~\ref{table:inter_Journal} presents the top 10 journals from which inter-collaborative papers were initially published and subsequently retracted. The majority of the published journals were of the Q1 category across various publishing houses. However, the highest number of retractions occurred in Q1 journals, with the majority of retractions originating from Hindawi publishers as per the top 10 journals. The highest retraction rate was observed in \textit{PLoS One}, accounting for 8.09\% of retractions, followed by \textit{Computational Intelligence and Neuroscience} with 4.5\% of retractions published by Hindawi.


\begin{table}[!h]
\caption{Intra-country: List of top 10 journals with number of retractions,  journal impact and publisher detail. }
\begin{tabular}{|l|c|c|c|l|}
\hline
\textbf{Journl Name} &
  \textbf{\begin{tabular}[c]{@{}c@{}}No of \\ Retractions\end{tabular}} &
  \textbf{In \%} &
  \textbf{Quartile} &
  \textbf{Publisher} \\ \hline
\begin{tabular}[c]{@{}l@{}}Journal of Ambient Intelligence and   \\ Humanized Computing\end{tabular}          & 335 & 12.94 & Q1          & Springer                   \\ \hline
Journal of Physics: Conference Series                                                                         & 268 & 10.35 & Q4          & IOP Publishing             \\ \hline
\begin{tabular}[c]{@{}l@{}}IOP Conference Series: \\ Materials Science and Engineering\end{tabular}           & 161 & 6.22  & Proceedings & IOP Publishing             \\ \hline
\begin{tabular}[c]{@{}l@{}}The International Journal of Electrical   \\ Engineering \& Education\end{tabular} & 33  & 1.27  & Q4          & SAGE Publications          \\ \hline
Cluster Computing                                                                                             & 31  & 1.2   & Q2          & Springer                   \\ \hline
\begin{tabular}[c]{@{}l@{}}International Journal of Mechanical \\ and Production Engineering Research \\ and Development\end{tabular} &
  31 &
  1.2 &
  No Impact &
  Trans Stellar \\ \hline
Wireless Personal Communications                                                                              & 27  & 1.04  & Q2          & Springer                   \\ \hline
PLoS One                                                                                                      & 25  & 0.97  & Q1          & PLoS                       \\ \hline
RSC Advances                                                                                                  & 23  & 0.89  & Q2          & Royal Society of Chemistry \\ \hline
The Journal of Biological Chemistry &
  21 &
  0.81 &
  Q4 &
  \begin{tabular}[c]{@{}l@{}}American Society for \\ Biochemistry and \\ Molecular Biology\end{tabular} \\ \hline
\end{tabular}
\label{table:intra_Journal}
\end{table}
\begin{table}[!h]
\caption{Inter-country: List of top 10 journals with number of retractions,  journal impact and publisher detail. }
\begin{tabular}{|l|c|c|c|l|}
\hline
\textbf{Journl Name} &
  \textbf{\begin{tabular}[c]{@{}c@{}}No of \\ Retractions\end{tabular}} &
  \textbf{In \%} &
  \textbf{Quartile} &
  \textbf{Publisher} \\ \hline
PLoS One                                      & 53 & 8.09 & Q1 & PLoS     \\ \hline
Computational Intelligence and Neuroscience   & 30 & 4.58 & Q1 & Hindawi  \\ \hline
\begin{tabular}[c]{@{}l@{}}Journal of Ambient Intelligence and   \\ Humanized Computing\end{tabular} &
  23 &
  3.51 &
  Q1 &
  Springer \\ \hline
Aggression and Violent Behavior               & 19 & 2.9  & Q1 & Elsevier \\ \hline
Arabian Journal for Science and   Engineering & 19 & 2.9  & Q1 & Springer \\ \hline
Journal of Healthcare Engineering             & 17 & 2.6  & Q3 & Hindawi  \\ \hline
Journal of Nanomaterials                      & 13 & 1.98 & Q1 & Hindawi  \\ \hline
\begin{tabular}[c]{@{}l@{}}Proceedings of the Institution of Mechanical Engineers, \\ Part E: Journal of Process Mechanical Engineering\end{tabular} &
  11 &
  1.68 &
  Q2 &
  SAGE Publications \\ \hline
Adsorption Science \& Technology              & 11 & 1.68 & Q2 & Hindawi  \\ \hline
Advances in Materials Science and Engineering & 10 & 1.53 & Q2 & Hindawi  \\ \hline
\end{tabular}
\label{table:inter_Journal}
\end{table}

\section{Country Collaboration Network}
\begin{figure}[!h]
    \centering
    \includegraphics[width=0.85\linewidth]{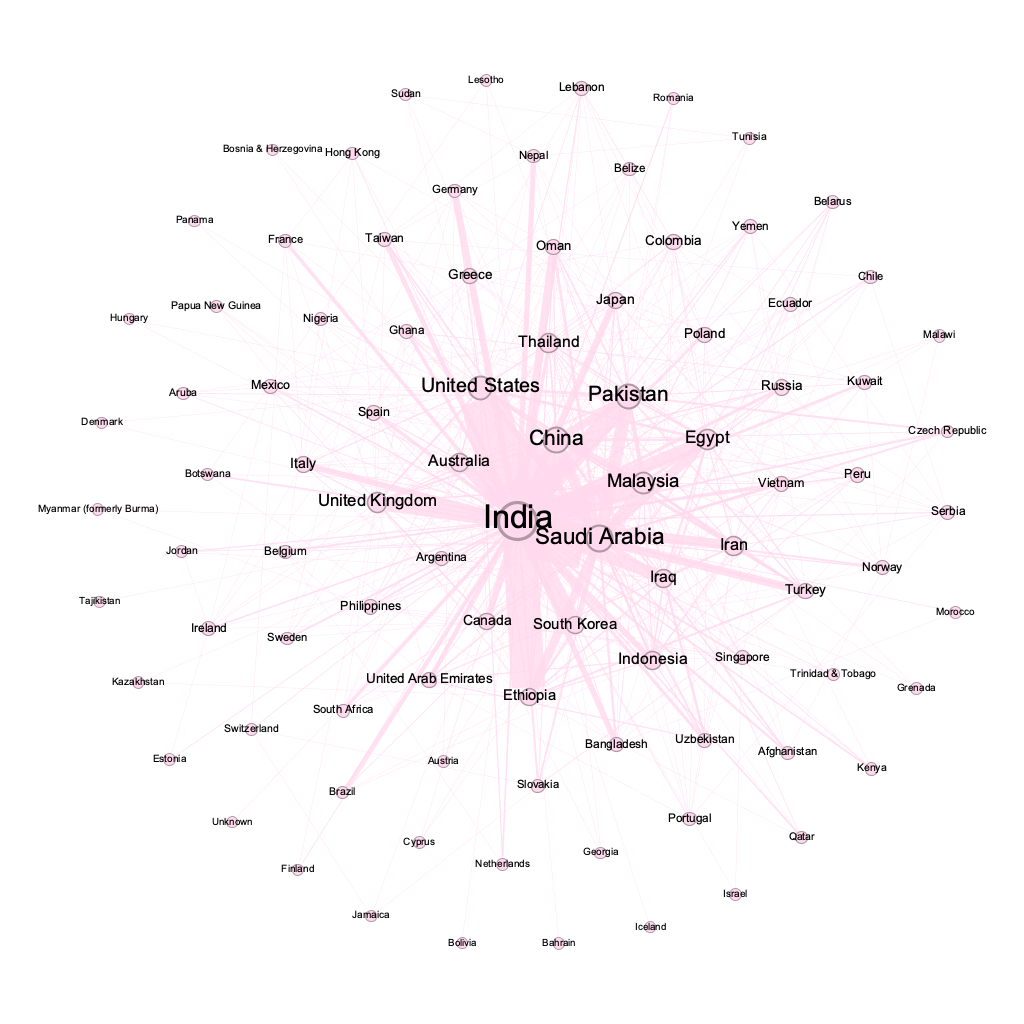}
\caption{Country collaboration network with 94 nodes and 561 unique edges.}
\label{fig:network}   
\end{figure}

Indian researchers were involved in 655 retracted papers, collaborating with researchers from 93 countries. Figure~\ref{fig:network} illustrates the country collaboration network, with India positioned as the central node, engaging in numerous collaborations, particularly with Saudi Arabia. The second most common collaboration occurred with the USA, followed by China, Ethiopia, and Malaysia, among others. Specifically, India collaborated with Saudi Arabia in 166 retracted papers (14.3\%), the USA in 117 (10.1\%), and China in 104 retracted papers (8.9\%). Moreover, collaborations with Ethiopia (83) and Malaysia (58) jointly contributed to 12\% of retractions. Additionally, notable collaborations were observed with Pakistan, South Korea, the United Kingdom, Egypt, Iraq, Australia, and Iran, highlighting India's global engagement in cases of misconduct.

Considering India's collaboration with 93 countries, a total of 1158 connections were formed. One-third (33.4\%) of retractions occurred in collaboration with the top three countries: Saudi Arabia, USA, and China, while 62\% of retractions originated from collaborations with the top 10 countries: Saudi Arabia, USA, China, Ethiopia, Malaysia, Pakistan, South Korea, the United Kingdom, Egypt, Iraq, Australia, and Iran.
%

\section{Conclusion}

The current investigation into retracted publications affiliated with India sheds light on valuable insights regarding the types of retractions and the institutions involved. This study analyzed 3244 retracted papers with Indian affiliations, among which Indian authors solely authored 2589, and 655 were collaborations with authors from 93 countries. The upward trend in the number of retractions and the decreasing time until retraction (measured in months) indicate that a majority of the retracted Indian articles were published after 2010. The primary reasons for retraction are fake peer review, followed by data integrity and plagiarism.

Upon examining intra- and inter-country collaboration patterns, it was observed that retractions of papers solely affiliated with India are mainly attributed to fake peer review, data integrity issues, and plagiarism. Plagiarism is primarily observed in papers published in conference proceedings or those lacking impact (not indexed in Scopus). Conversely, papers retracted with international collaboration were retracted due to various reasons.

Several private institutes and universities offer financial incentives for each publication, potentially leading to a focus on quantity over quality and encouraging unethical publishing practices. Furthermore, the types of Indian institutions and universities involved in such retractions were analyzed. It was discovered that 60\% of retractions originated from private universities, institutes, or colleges, while the remaining retractions were from public universities, national institutes, research centers, or a small fraction from medical colleges or hospitals. The primary cause of retraction from these institutions was examined, revealing that fake peer review is the main cause in private institutes or colleges, whereas data integrity issues are more prominent in public institutes or universities and medical colleges.

Notably, prominent publishers such as Springer, IOP Publishing, and Elsevier account for more than half of the retractions. Overall, intra-country collaborations account for approximately half of the total retractions from the top two journal quartiles (Q1 and Q2), while in inter-country collaborations, roughly one-third of retractions originate from the top two quartiles. Additionally, 45.6\% of retractions originated from the top five collaborators: Saudi Arabia, USA, China, Ethiopia, and Malaysia.

In essence, collaboration alone cannot guarantee the quality of research outcomes. It is crucial to address related actions about the researcher's adherence to scientific norms within the scientific community. Scientific misconduct closely correlates with individual awareness of scientific integrity, potentially undermining both collaboration networks and scientific integrity as highlighted by Sharma et al. (2023).

For institutions with higher retraction rates, there is a pressing need to prioritize the production of high-quality research and enhance scientific ethics through integrity training. To mitigate the adverse impact of recurrent misconduct among collaborators, it is advisable to establish a comprehensive and legally protected system for reporting scientific misconduct, along with mentorship programs for institutions or research groups~\citep{lu2013retraction, davis2007causal, kornfeld2012perspective}.

Moreover, researchers themselves need to recognize the repercussions of scientific misconduct on their careers and the integrity of national scientific endeavors~\citep{shahare2020historicizing}. Private institutes ought to alleviate the pressure of "publish or perish" linked to annual performance evaluations, thereby reducing the temptation to engage in fraudulent practices such as fake peer review for expedited publications or the fabrication of data to produce extensive papers. Researchers must receive training to recognize the factors contributing to plagiarism and understand its consequences. Additionally, it is vital to establish appropriate policies for governing national science, particularly those addressing issues such as fake peer review, plagiarism, and the clarification of authorship contributions. These policies facilitate transparency regarding individual research contributions, as advocated by COPE (2020).

The study has several limitations regarding data availability for addressing certain questions. Firstly, the information regarding corresponding authors is not provided, and it is unclear which author is affiliated with which institute. Consequently, determining who is responsible for the false research and who the primary contributor (first author) is becomes challenging. Additionally, the author names are abbreviated, making it difficult to discern the authors' genders. Thus, we cannot ascertain the gender distribution in the retractions or their roles in the retracted work. Similarly, information about the types of institutions the authors belong to is also lacking.

\section*{Data availability statement}
The data is freely available on Crossref (\url{https://www.crossref.org/}).

\section*{Conflict of Interest}
The author declares no conflict of interest.

\printcredits
\bibliographystyle{cas-model2-names}

\bibliography{cas-refs}
\end{document}